**Title:**

**Modeling membrane morphological change during autophagosome formation**

Yuji Sakai[1,2*], Ikuko Koyama-Honda[1], Masashi Tachikawa[2,3], Roland L. Knorr[1,4,5], and Noboru Mizushima[1,6*]

[1]Department of Biochemistry and Molecular Biology, Graduate School of Medicine, The University of Tokyo, Bunkyo-ku, Tokyo 113-0033, Japan

[2]Interdisciplinary Theoretical and Mathematical Sciences (iTHEMS) Program, RIKEN, Wako, Saitama 351-0198, Japan

[3]Institute for Frontier Life and Medical Sciences, Kyoto University, Shogoin Kawahara-cho, Sakyo-ku, Kyoto 606-8507 Japan

[4]Max Planck Institute of Colloids and Interfaces, Department of Theory & Bio-Systems, 14424 Potsdam, Germany

[5]Max-Planck-Institute of Molecular Plant Physiology, Dept. Molecular Physiology, 14424 Potsdam, Germany

[6]Lead Contact

*Correspondence: yuji-sakai@m.u-tokyo.ac.jp and nmizu@m.utokyo.ac.jp.

**Summary**

Autophagy is an intracellular degradation process that is mediated by *de novo* formation of autophagosomes. Autophagosome formation involves dynamic morphological changes; a disk-shaped membrane cisterna grows, bends to become a cup-shaped structure, and finally develops into a spherical autophagosome. We have constructed a theoretical model that integrates the membrane morphological change and entropic partitioning of putative curvature generators, which we have used to investigate the autophagosome formation process quantitatively. We show that the membrane curvature and the distribution of the curvature generators stabilize disk- and cup-shaped intermediate structures during autophagosome formation, which is quantitatively consistent with *in vivo* observations. These results suggest that various autophagy proteins with membrane curvature-sensing properties control morphological change by stabilizing these intermediate structures. Our model provides a framework for understanding autophagosome formation.

**Introduction**

Membrane-bound organelles compartmentalize eukaryotic cells and adopt various characteristic shapes such as disk-shaped cisternae, tubules, spherical vesicles, and their intermediate structures. Organellar morphology is regulated by protein and lipid compositions. Because organellar shape directly relates to organellar function, it is important to understand the mechanisms regulating organelle morphology.

All intracellular membranes are in a fluid state. Because membranes are extremely thin (approximately 5-nm thick), they can be considered quasi-two-dimensional surfaces with morphologies that are governed primarily by two elastic parameters, bending elasticity and spontaneous curvature (Helfrich, 1973). Bending elasticity corresponds to the strength of the membrane to resist bending, and spontaneous curvature corresponds to the curvature that an unconstrained membrane would adopt. Because quantitative comparisons between experimentally observed and theoretically calculated shapes have confirmed mesoscopic descriptions, the mechanisms by which intracellular force-generating machines drive remodeling of membrane-bound organelles have been explored (Bassereau et al., 2018; Lipowsky, 1991; Ramakrishnan et al., 2018).

Macroautophagy, simply referred to as autophagy hereafter, is a membrane-mediated bulk degradation process. In this process, a portion of the cytoplasm is engulfed by an organelle termed the autophagosome, which has a spherical double-membrane structure with a typical diameter of 0.5–1.0 μm (Mizushima et al., 2011). Autophagosomes then fuse with lysosomes to degrade the engulfed materials. Autophagosomes are generated *de novo* in the cytoplasm by the expansion of disk-shaped membrane cisternae termed phagophores (or isolation membranes). As they increase in size, the morphologies of phagophores change from disk- to cup-shaped. Eventually, they develop into double-membraned spherical structures (Figure 1).

Although approaches from molecular biology and genetics have been introduced to the study of this subject, the mechanism underlying these characteristic morphological changes remains largely unknown (Mizushima, 2018). In the meantime, physical and mathematical modeling approaches have begun to be used. The phagophore has three distinct regions with different membrane curvatures: two closely juxtaposed sheets, the inner (*a* in Figure 1) and outer (*b* in Figure 1) membranes, and the connecting rim (*c* in Figure 1). The distance between the outer and inner membranes is considered to be less than 30 nm, with a rim that is not swollen (note that the intermembrane space, which is often observed widened by electron microscopy, is actually very thin if it is appropriately fixed) (Hayashi-Nishino et al., 2009; Uemura et al., 2014; Ylä-Anttila et al., 2009), making the rim highly curved and energetically expensive (Nguyen et al., 2017). The bending energy minimization principle (Seifert et al., 1991) indicates that a disk-shaped membrane that is larger than a certain area should bend into a closed spherical structure (stomatocyte) so as to reduce the bending energy at the rim. The rim energy is proportional to the rim length, which is shortened by disk bending. This mechanism was recently proposed to be important for understanding autophagosome formation (Knorr et al., 2012) and the curling of the Golgi cisternae (Campelo et al., 2017). However, the process of autophagosome formation *in vivo* does not fit this simple model in two respects. First, the model predicts that the autophagosomes would be considerably smaller; a disk will be closed when its radius is only five times larger than that of the rim ($r$). If the length of the intermembrane space ($2r$) is <30 nm, the final size of an autophagosome ($2R$) would be <75 nm, which is much smaller than regular autophagosomes, which are 0.5–1.0 μm. Second, because intermediate cup-shaped structures are energetically unstable (Knorr et al., 2012), they would abruptly transform into spherical structures, which is inconsistent with *in vivo* observations of cup-shaped structures existing for several minutes (Tsuboyama et al., 2016). The first issue can be resolved because if the rim curvature is stabilized by its spontaneous curvature, the disk shape will be more stable, and thus, autophagosomes can become larger (Knorr et al., 2012). However, regarding the second issue, the mechanism behind the unexpected stability of the intermediate structures remains unknown. Thus, it is desired to understand the dynamics of the continuous membrane morphological transition during the whole process of autophagosome formation and to reveal the mechanism regulating the membrane curvature, particularly in the intermediate structures.

In this study, we investigated the dynamics of the continuous membrane morphological transition during autophagosome formation. We hypothesized that the morphological change is governed by the spatiotemporal regulation of putative curvature generators. Curvature generators can be various biomolecules, including proteins with partially inserted or wedge-shaped membrane domains and conical lipids (McMahon and Boucrot, 2015). Indeed, many autophagy-related (ATG) proteins contain curvature-generating (or sensing) domains (Nguyen et al., 2017). Heterogeneous distributions of autophagosomal proteins may regulate membrane curvature and stabilize autophagosome intermediates. The morphology is determined by minimizing the total energy, comprising both the bending energy of the phagophore and the partitioning entropic energy of the curvature generators. We show that this feedback effect between the membrane shape and the distribution of curvature generators can stabilize all the structures appearing during autophagosome formation: disks, cup-shaped intermediates, and spherical structures. The results obtained from this model are quantitatively consistent with observations of *in vivo* experiments. Furthermore, our model predicts that curvature generator abundance is positively correlated with the resulting autophagosome size, which should be helpful for identifying curvature generators experimentally.

**Results**

**Modeling phagophore morphological changes**

Here, we present a model to evaluate the membrane morphological change and the distribution of curvature generators during autophagosome formation. The former is captured by the elastic bending energy of the membrane, $F_{bend}$, and the latter is captured by the partitioning entropic energy of curvature generators, $F_{part}$. The total free energy of a membrane structure, $F_{tot}$, consists of the two parts:

$$F_{tot} = F_{bend} + F_{part}. \tag{1}$$

We assume that the membrane morphology and the distribution of curvature generators are equilibrated and can be obtained by minimizing the total free energy for a given membrane area $A$ and area $A_\emptyset$ occupied by the curvature generators. Hereafter, the area of curvature generators, $A_\emptyset$, is referred to as the abundance. Specifically, the mechanical relaxation time of biological membranes is $\tau_{mech} = 1\ ms$ (Campelo et al., 2017), whereas the time required for autophagosome formation is $\tau_{flux} = 10\ min$ (Tsuboyama et al., 2016). Because $\tau_{mech} \ll \tau_{flux}$, the system should be mechanically equilibrated.

The elastic bending energy of a membrane is associated with the deviation of total curvature from the spontaneous curvature over the entire membrane and is proportional to the bending rigidity of the membrane,

$$F_{bend} = \sum_{i=\pm,r} \frac{\kappa}{2} \int (J_i - \overline{J_i})^2 dA_i, \tag{2}$$

where $\kappa$ is the rigidity and $J_i$, $\overline{J_i}$, and $A_i$ are the total curvature, spontaneous curvature, and area, respectively, of region $i$ (Helfrich, 1973). The geometry of the phagophore consists of three distinct regions with different curvatures (Figure 2), and the summation over index $i$ includes the outer ($+$), inner ($-$), and rim ($r$) membranes. Based on our previous observations (Tsuboyama et al., 2016), we modeled the geometry of the outer and inner membranes as part of an ellipsoid structure with a bending angle $\alpha$ (Figure 2). The distance between the outer and inner membranes ($2r$) is much smaller than that of the lateral dimension ($\sim 2R$) and is assumed to be constant based on observations made via electron microscopy (Hayashi-Nishino et al., 2009; Uemura et al., 2014; Ylä-Anttila et al., 2009). There should be some mechanism to keep the rim radius constant but that is not known. The rim geometry is modeled as part of a torus with a fixed minor radius $r = 10\ nm$. The total curvature $J_i$ and the surface element $dA_i$ of the region $i$ in Equation (2) are determined from the geometric parameters, such as the bending angle $\alpha$, the lateral radius $R$, the aspect ratio $\gamma$, and the rim radius $r$. The complete definitions of total curvature and surface element are given in the Supplemental Information.

We consider the curvature generators distributed on the membrane. The curvature generators induce spontaneous membrane curvature and stabilize highly curved regions by decreasing the bending energy. Proteins that are partially inserted into membranes, such as those containing amphipathic helixes, create various curvatures in a concentration-dependent manner (Campelo et al., 2008). The membrane spontaneous curvature $\bar{J}_i$ in Equation (2) is assumed to be proportional to the area fraction $\phi_i$ of the curvature generators at region $i$,

$$\bar{J}_i = \frac{\zeta}{2}\phi_i, \tag{3}$$

where $\zeta$ is a proportionality coefficient (Campelo et al., 2008). The area fraction of curvature generators is uniform in each region. The curvature generators possess the partitioning entropic energy

$$F_{part} = -k_B T \sum_{i=\pm,r} \left(\phi_i ln\phi_i + (1-\phi_i)ln(1-\phi_i)\right)\frac{A_i}{a_\phi}, \tag{4}$$

where $k_B$, $T$, and $a_\phi$ are the Boltzmann constant, temperature, and surface area of a curvature generator, respectively. The surface area $A_i$ of the region $i$ is given in the Supplemental Information. Maldistribution of curvature generators increases the partitioning entropic energy. The balance between the bending energy $F_{bend}$ and the partitioning entropic energy $F_{part}$ determines the membrane morphology and distribution of curvature generators. The model parameters are summarized in Table 1.

**Stabilized morphology predicted by the model**

First, we determined the change in membrane morphology of a thin cisterna as the area increases. Figure 3 shows a cross-section of stable membrane structures for different total areas $A$ in the cases with no curvature generators (pannel A) and with the fixed curvature generators $A_\phi = 0.1\ \mu m^2$ (panel B). The line colors indicate the area fraction of the curvature generators in each region. The membrane morphology and the distribution of curvature generators are obtained by minimizing the free energy $F_{tot}$ for each membrane area $A$. In the absence of curvature generators, a disk shaped-structure becomes a closed structure at $A = 0.001\ \mu m^2$, and there is no stable cup-shaped intermediate structure (Figure S1). In the presence of curvature generators, a disk-shaped structure stays stable at $A < 2.5\ \mu m^2$ and then transforms into a cup-shaped structure at $A = 2.5\ \mu m^2$, which continuously bends and is eventually closed at $A = 3.1\ \mu m^2$. This means that the curvature generators make the disk size larger, as predicted previously (Knorr et al., 2012), while stabilizing the intermediate cup shape. The diameter of typical autophagosomes is 0.5–1.0 μm, corresponding to an area of 2–10 μm². This suggests the necessity of curvature generators in stabilizing the phagophore morphology observed *in vivo*.

The stability of each membrane shape is analyzed from the potential energy surface. Figure 4*A* shows the total free energy as a function of the bending angle α for different membrane areas $A$ with a fixed abundance of curvature generators. The potential is minimized when $\alpha = 0$ and $\alpha \approx \pi$ for small $A$ (blue curve) and large $A$ (green curve), respectively. This means the disk and the closed shape are stable for small and large $A$, respectively. The potential has double minima when $\alpha = 0$ and $\alpha \approx 0.4\pi$ for intermediate $A$ (red curve). The minimum energy levels correspond to a disk- and cup-shaped structure, and they are separated by the potential barrier with $\sim k_B T$ (Figure 4*A*, inset), which is comparable to the thermal fluctuation.

The continuous membrane morphological change is calculated with increasing membrane area. Figure 4*B* shows the bending angle $\alpha$ and the aspect ratio $\gamma$ as the membrane area $A$ increases for fixed $A_\phi = 0.1, 0.15$, and $0.2\ \mu m^2$. For all $A_\phi$, the bending angle $\alpha$ remains 0 until the area reaches a critical size ($A_d$, the maximum disk area) and jumps to $\alpha \approx 0.5\pi$ at $A_d$. Then, it increases gradually and finally reaches $\pi$. This represents a membrane morphological change from a disk ($\alpha = 0$) to a cup-shaped vesicle ($0 < \alpha < \pi$), and eventually to a closed structure ($\alpha \cong \pi$). The maximum disk area $A_d$ increases with $A_\phi$ because a larger stable disk shape requires more curvature generators in order to stabilize the curvature of its larger rim region.

The aspect ratio $\gamma$ takes the maximum value immediately after the start of bending at $A_d$. Thus, the cup-shaped structures are elliptically deformed during intermediate stages. As the membrane area $A > A_d$ is increased further, $\alpha$ gradually increases, whereas $\gamma$ gradually decreases. Under such conditions, the elliptic intermediates become spherical. Considering a cup-shaped structure, an elongated (elliptic) spheroid has a smaller rim area and is more stable compared with a complete sphere. However, when the rim area becomes very small immediately before closure, an elongated spheroid has a larger curved surface and is less stable than a sphere. Thus, there is a transition from the elliptic cup-shape to the complete sphere. This tendency does not depend on the abundance of curvature generators. Intermediate cup-shaped structures are also stabilized for $A > A_d$ even for spherical geometry with fixed $\gamma = 1$ (Figure S2).

**Stabilization of autophagosomal intermediates by redistribution of curvature generators**

Next, we determined the redistribution of curvature generators as the membrane area increases. Figure 5*A* shows the average total curvature $\langle J_i \rangle = \int J_i dA_i /A_i$ of the region $i$, while Figure 5*B* shows the spontaneous curvature $\bar{J}_i$ generated by the curvature generators for curvature generators with $A_\phi = 0.1\ \mu m^2$. For a small membrane area $A < A_d \cong 2.5 \mu m^2$, where the membrane adopts a disk shape, the rim has a very high average curvature ($\langle J_r \rangle \sim 1/r$), while the flat part (later becoming the outer and inner membranes) has zero curvature ($\langle J_\pm \rangle \sim 0$). The fraction of area with curvature generators on the rim $\phi_r$ is much higher than $\phi_\pm$ while $\phi_+ = \phi_-$. The rim spontaneous curvature is comparable to the average curvature, $\bar{J}_r \sim \langle J_r \rangle$ (with a deviation of only a few percent), and then stabilizes the rim curvature. As the membrane area increases, the rim area fraction $\phi_r$ decreases and the difference between the average curvature and the spontaneous curvature increases. The deviation becomes approximately 10% at $A = A_d$. The deviation raises the rim bending energy and drives the morphological transition to the cup shape.

To stabilize a highly curved rim, the area fraction of the curvature generator on the rim $\phi_r$ must be twenty times higher than $\phi_\pm$. This biased distribution of the curvature generators raises the partitioning entropic energy. The entropic effect increases with membrane area (Figure S3*A*). There are two competitive flows of curvature generators; one is the entropy-induced flow from the rim to the inner and outer membranes, and the other is the curvature-induced flow in the opposite direction for rim stabilization. For $A < A_d$, the curvature-induced flow surpasses the entropy-induced flow. However, as the area increases, the entropic effect becomes stronger, and the entropy-induced flow overcomes the curvature-induced flow for $A \geq A_d$ (Figure S3*B*). The curvature generators flow from the rim to the outer membrane. If all the curvature generators at the rim move to the outer membrane when the transition occurs, the outer spontaneous curvature becomes higher than the membrane total curvature, which increases the bending energy. Thus, some amount of curvature generators stay at the rim, and the distribution of curvature generators settles down to a proper value.

For $A > A_d$, the outer and inner membranes have positive and negative curvatures, respectively, leading to an asymmetric distribution of the curvature generators. The asymmetric distribution induces a spontaneous curvature difference between the outer and inner membranes and stabilizes the cup-shaped intermediate structures. A decrease in the rim area, along with morphological transition, maintains the mobilization of the curvature generators from the rim to the outer membrane and induces a further increase in the spontaneous curvature and bending of the outer membrane.

**Controls of the autophagosome size by curvature generators**

Figure 6 shows the heat map of the bending angle $\alpha$ in two-dimensional parameter space with $A_\phi$, the abundance of curvature generators, and $A$, the total membrane area, as $x$- and $y$-axes, respectively. The heat map is divided into three regions by the boundary formed by $A = A_d$ and $A = A_s$ where the boundary $A = A_s$ is defined at $\alpha = 0.97\pi$. For $A < A_d$, $\alpha = 0$ and the membrane takes a disk shape. For $A_d < A < A_s$, the membrane takes a cup shape ($0 < \alpha < \pi$), where $\alpha$ increases with $A$ and the membrane is gradually closing. The boundary $A = A_d$ is the first-order shape transition line. For $A > A_s$ with $\alpha \approx \pi$, the membrane takes a closed structure. The transition from a cup to a closed shape is continuous.

The phase boundaries $A_d$ and $A_s$ linearly increase with the area (i.e., the abundance) of curvature generators, $A_\phi$. This indicates that the maximum disk area $A_d$ and the closed area $A_s$ are positively correlated with each other, suggesting that both sizes are controlled by the abundance of curvature generators. Indeed, $A_d$ and $A_s$ demonstrate a positive linear correlation $A_s \cong 1.3 A_d$ (see Figure 7C, dotted line).

**Regulation of the abundance of curvature generators**

The abundance of the curvature generators $A_\phi$ has been assumed to be fixed so far, but it can be variable during autophagosome formation. Curvature generators can be dynamically recruited from the cytosol. When the cytosol has a sufficient abundance of curvature generators and the exchange between the membrane and the cytosol is rapid enough, their abundance is equilibrated. The abundance $A_\phi$ is obtained by minimizing the grand potential

$$\Omega = F_{tot} - \int \mu \phi_i \, dA_i / a_\phi, \tag{5}$$

for a given total membrane area $A$. The chemical potential $\mu$ defines the binding affinity of the curvature generators to the membrane and may increase with the membrane curvature such that $\mu \propto J$.

White arrows in Figure 6 indicate the relationship between the abundance of the curvature generators $A_\phi$ and the membrane area $A$ when the chemical potential $\mu$ is constant (dashed line) or proportional to the membrane curvature (dotted line). If the chemical potential $\mu$ is positively constant, the abundance of the curvature generator, $A_\phi$, increases with the membrane area $A$. According to the phase diagram in Figure 6, only the disk-shape is stable and neither the cup- nor the closed-shape appears because the abundance of curvature generators is sufficient to stabilize the rim of the disk (Figure 6, the dashed line). This is also the case if the chemical potential is proportional to the membrane curvature (Figure 6, dotted line). These results are not consistent with *in vivo* observations (Figure 6, solid line), where the change in GFP-ATG2A intensity is shown (see the next section). These results suggest that the abundance of curvature generators is not simply determined by the equilibrium between the membrane and the cytosol. The supply of curvature generators may be limited or controlled by outside factors. Indeed, it is known that the amount of many ATG proteins increases during membrane elongation but decreases before closure (Honda et al. 2013). The binding affinity of the curvature generators may also be regulated by the post-translational modification of membrane molecules, such as phosphorylation by ULK1 and dephosphorylation of PI3P.

**Morphological changes during *in vivo* autophagosome formation**
We compare morpho-dynamics between our model and *in vivo* experimental observations. Figure 7*A* shows time-lapse frames of live-cell imaging of autophagosome formation in starved mouse embryonic fibroblasts (MEFs) (Video S1). Microtubule-associated protein 1 light chain 3B (LC3B), one of the Atg8 homologs in mammals, is conjugated to the lipid phosphatidylethanolamine in autophagosomal membranes, and mRuby3-fused LC3B (mRuby3-LC3B) is uniformly distributed on the autophagosomal membrane (Kabeya et al., 2000; Tsuboyama et al., 2016). Among ATG proteins, we found that green fluorescent protein-fused ATG2A (GFP-ATG2A) is present primarily on the highly curved rim of autophagosome intermediates (Figure. 7A). Although GFP-ATG2A is overexpressed, this localization pattern does not change when it is expressed at a level comparable to or even lower than that of endogenous ATG2A (data not shown).
The abundance of autophagy-related proteins on autophagosomes is regulated temporally during autophagosome formation (Koyama-Honda et al., 2013; Suzuki et al., 2013). As previously reported (Koyama-Honda et al., 2013), the intensity of the mRuby3-LC3B signal increases and plateaus during autophagosome formation (Figure S4, S5). In contrast, the GFP-ATG2A signal intensity increases, reaches a peak, and decreases (Figure S4, S5). Although a fixed abundance of curvature generators is used in the above theoretical analysis, it is logical to consider that this changes along with membrane growth as the abundance of many ATG proteins also change (Koyama-Honda et al., 2013). Because of the similarity in the behaviors between the observed GFP-ATG2A signal and supposed curvature generators, it is assumed that the intensities of mRuby-LC3B and GFP-ATG2A are proportional to the total membrane area $A$ and the area of the rim (and thus the abundance of curvature generators $A_\phi$) based on their localization (Figure 7*A*). The intensities of mRuby-LC3B and GFP-ATG2A are positively correlated during the initial phase, when they increase (Figure S6). The correlation can be fitted with a second polynomial and obtained as

$$\bar{A}_\phi = a\bar{A} + (1-a)\bar{A}^2, \quad (6)$$

with $a = 1.5$, $\bar{A}_\phi = A_\phi/A_\phi^0$, and $\bar{A} = A/A^0$, where $A_\phi^0$ and $A^0$ are each normalization factors. The case with $A_\phi^0 = 0.3 \, \mu m^2$ and $A^0 = 10 \, \mu m^2$ is shown as the *in vivo* curve in Figure 6 and solid curve in Figure 7*B*.

Figure 7*B* shows a comparison of the bending angle α between the experiment and the above model results. The dots show experimental data extracted from seven independent autophagosomes *in vivo*. For each step during autophagosome formation, the membrane area $A$ and the bending angle $\alpha$ are extracted by fitting the fluorescence images with part of an ellipsoid (Figure S7). The solid line in Figure 7B shows the model result with $A_\phi^0 = 0.3 \, \mu m^2$ and $A^0 = 10 \, \mu m^2$. The experimental data shows that the morphological transition from a disk ($\alpha \cong 0$) to a cup-shaped structure ($\alpha > 0$) occurs at $A_d \cong 0.6 A_s$, after which the cup-shaped structure is gradually closed ($\alpha \cong \pi$) at $A_s$. The model indeed shows that the morphological transition occurs at $A_d \cong 0.6 A_s$ and a cup-shaped structure is continuously closed thereafter. Therefore, our model quantitatively predicts the experimental data.

**Size scaling law of the phagophore and autophagosome**

As shown in Figure 6, our model suggests that the abundance of the curvature generators regulates both the maximum disk area $A_d$ and the closed spherical area $A_s$, as these areas are well correlated. Figure 7*C* shows the relationship between $A_d$ and $A_s$ experimentally taken from 89 autophagic structures in starved MEFs (red dots), with a positive correlation between $A_d$ and $A_s$. The quantitative relationship between the two variables can be expressed as $A_s \cong 1.8 A_d$, which has a slightly higher slope than that obtained from the model with a fixed abundance of curvature generators, that is, $A_s \cong 1.3 A_d$ (shown as the dotted line in Figure 7*C*). This discrepancy can be resolved if the abundance of curvature generators changes during autophagosome formation according to the fitting $\bar{A}_\phi$. With the normalization factor $A^0 = 40 A_\phi^0$, the slope becomes steeper (Figure 7*C*, solid line) and consistent with the experimental result, that is, $A_s \cong 1.8 A_d$.

**Effects of two different types of curvature generators**

Thus far, the distribution of a single curvature generator has been considered for simplicity. However, multiple autophagy-related proteins are known to be present on autophagosomal membranes. Notably, most of them (e.g., ATG12, ATG5, and ATG16L1) are present on the outer autophagosomal membrane rather than the inner membrane and the rim (Mizushima et al., 2011). In addition, the autophagic membrane contains negative curvature generators such as phosphatidylethanolamine, which has a negative spontaneous curvature (Kamal et al., 2009). These proteins and lipid compositions could also produce spontaneous curvature. Thus, two types of curvature generators with different spontaneous curvatures are considered in the model. The effects of the spontaneous curvature and partitioning energy of two different types of curvature generators are incorporated into the model (Supplemental Information). Curvature generators with a positive (negative) spontaneous curvature are referred to as positive (negative) curvature generators. Here, the curvature generator used in the previous model (Figure 4B) is referred to as a strongly positive curvature generator. The addition of weakly positive curvature generators stabilizes both disk- and cup-shaped structures. The morphological transition in this situation is similar to the case with only the strongly positive curvature generator (Figure 8A, red curve). The weakly positive curvature generators are distributed uniformly on the membrane, while strongly positive curvature generators are localized mainly to the highly curved rim (Figure 8B). Negative curvature generators stabilize the negatively curved inner membrane of cup-shaped structures, but they cannot stabilize disk-shaped structures or the positively curved rim. Thus, a disk starts bending at a smaller A and becomes a smaller spherical structure (blue curve in Figure 8A). The negative spontaneous curvatures tend to be distributed on the inner membrane but not the positively curved rim (Figure 8C).

**Discussion**

We hypothesized that morphological change in the phagophore membrane is governed by the spatiotemporal regulation of putative curvature generators. The morphology is determined by minimizing the total energy, which comprises the bending energy of the membrane and the partitioning entropic energy of the curvature generators. In order to investigate how curvature generators regulate and stabilize the overall morphology of autophagosome intermediates, we constructed a simple model considering changes in phagophore morphology and partitioning of curvature generators. This model allows the curvature generators to localize at the highly curved rim and stabilize it. As a result, the disk shape is stabilized until a certain membrane size is reached. When the disk size exceeds a critical threshold value, the partitioning effect causes a shape transition from a disk to a spherical stomatocyte. At the onset of the transition, the curvature symmetry between the inner and outer membranes is broken, which is accompanied by an asymmetric distribution of the curvature generators, thus stabilizing cup-shaped intermediate structures. At this point, the majority of curvature generators are still present at the rim (Figure 5). Accordingly, we conclude that intermediate structures that appear during autophagosome formation can be stable at each time point. Furthermore, our model predicts that the abundance of curvature generators is positively correlated with the resulting autophagosome size. This prediction would help identify the actual curvature generators experimentally in future research.

The exact identity of the curvature generators for autophagosomes remains unknown. The shapes of autophagosomes could be spatiotemporally regulated by proteins and lipids (Nguyen et al., 2017). Many autophagy-related proteins demonstrate characteristic spatiotemporal distributions during autophagosome formation and contain curvature-sensing domains (Nguyen et al., 2017). In yeast cells, Atg2, Atg9, and Atg18 localize to the edge, whereas Atg1, Atg8, and the Atg12–Atg5–Atg16 complex are present on the surface membrane (Suzuki et al., 2013). In contrast to mammalian ATG2, yeast Atg2 localizes to two or three punctate structures at the edge of forming autophagosomes, which correspond to the contact sites between the phagophore and ER exit sites (Graef et al., 2013; Suzuki et al., 2013). Thus, yeast Atg2 may not be involved in rim stabilization. In contrast, yeast Atg20 and Snx4/Atg24 have BAR domains that are capable of inducing membrane curvature (Popelka et al., 2017). Although mammalian ATGs do not contain functional counterparts of Atg20 and Snx4/Atg24, these molecules can be candidate autophagosomal curvature generators, at least in yeasts. Indeed, yeast cells lacking both Atg20 and Snx4/Atg24 produce smaller autophagosomes and are defective in selective autophagy (Zhao et al., 2016). In mammalian cells, most ATG proteins are present on the convex-facing outer membrane, LC3/GABARAP family proteins are present on both outer and inner membranes, and DFCP1 is present near the rim of growing phagophores (Axe et al., 2008; Koyama-Honda et al., 2013; Mizushima et al., 2011). In the present study, we found that ATG2A is present on the rim in mammalian cells as well. Although we have not yet obtained conclusive evidence, ATG2 could be a candidate autophagosomal curvature generator based on the following features. ATG2 has a conserved amphipathic helix region that is required for localization to autophagosomes and lipid droplets (Chowdhury et al., 2018; Kotani et al., 2018; Tamura et al., 2017; Zheng et al., 2017). Cells with reduced ATG2A/B levels produce small autophagosome-like structures (Kishi-Itakura et al., 2014). Recently, structural and biochemical evidence has suggested that ATG2 has ER-to-autophagosome lipid-transfer activity (Maeda et al., 2019; Osawa et al., 2019; Valverde et al., 2019), which might also contribute to the regulation of autophagosomal size. However, the lipid transfer rate of ATG2 appears to be very slow (~0.017 lipid/s) (Maeda et al., 2019), which is much slower than the lipid mechanical relaxation time (Campelo et al., 2017). Recently it was proposed that ATG2 may have a faster lipid transfer rate (~100 lipid/s). Therefore, the membrane shape should be equilibrated at each time point during autophagosome formation irrespective of its expansion rate. However, another recent report suggests that the lipid transfer activity of ATG2 is much higher than the experimental estimate, ~100 lipid/s (Bülow and Hummer, 2020), which is comparable to the lipid mechanical relaxation time and could produce non-equilibrium effects. Even in this case, a mechanism to stabilize the shape of intermediate structures is required under non-equilibrium fluctuation. Lipid transfer from the ER to the outer leaflet of autophagosomal membranes may give rise to an asymmetric distribution of lipid molecules between the outer and inner leaflets. Such an asymmetry could induce a prolate deformation of a vesicle (Seifert et al., 1991). However, this is not observed during the formation of typical autophagosomes. The asymmetry may be resolved by putative lipid scrambling activity. In any case, it is important to differentiate a curvature-stabilizing effect from lipid-transfer activity in expanding the autophagosomal membrane.

**Limitations of the Study**
In this study, we investigated the effect of putative curvature generators on morphological change in membranes during autophagosome formation. A limitation of this study is that we have not yet identified the autophagosomal curvature generator(s) experimentally. Although ATG2 is a good candidate because it is present primarily at the highly curved rim and has an amphipathic helix region, we are still working to determine whether this is a primary curvature generator. Another remaining question is how the abundance of these curvature generators on the autophagosomal membrane is regulated. In this study, we propose that it is not simply determined by membrane morphology and the distribution of the curvature generators. We speculate that it is regulated by biochemical mechanisms such as protein phosphorylation rather than a physical mechanism. In our model, the membrane morphology is considered as only a part of an ellipse. It would be valuable to extend the model to deal with various shapes. We also assume the membrane morphology and the distribution of curvature generators are equilibrated at a given membrane area. It would be valuable to consider lipid transfer activity during membrane elongation.

## Methods

All methods can be found in the accompanying Transparent Methods supplemental file.

## Acknowledgements

We thank Drs. Atsushi Mochizuki and Takashi Taniguchi (Kyoto University) for helpful discussions. This work was supported by a Research Fellowship for Young Scientists (grant number JP18K14693) from the Japan Society for the Promotion of Science (JSPS) (to Y.S.) and Grants-in-Aid for Scientific Research (18H04874) from JSPS (to M.T.) and Exploratory Research for Advanced Technology (ERATO; grant number JPMJER1702) from the Japan Science and Technology Agency (JST) (to N.M.). The numerical computations have been performed with the RIKEN supercomputer system (HOKUSAI).

## Author Contributions

Conceptualization, Y.S., I.K-H., M.T., R.L.K., and N.M.; Methodology, Y.S., I.K-H., M.T., and N.M.; Investigation, Y.S., I.K-H., and N.M.; Formal analysis, Y.S. and I.K-H.; Writing, Y.S. and N.M.; Visualization, Y.S. and I. K-H; Funding acquisition, Y.S., I.K-H., M.T., and N.M.; Resources, Y.S., I.K-H., and N.M.; Supervision, Y.S. and N.M.

## Declaration of Interests

The authors declare no competing interests.

**Figures**

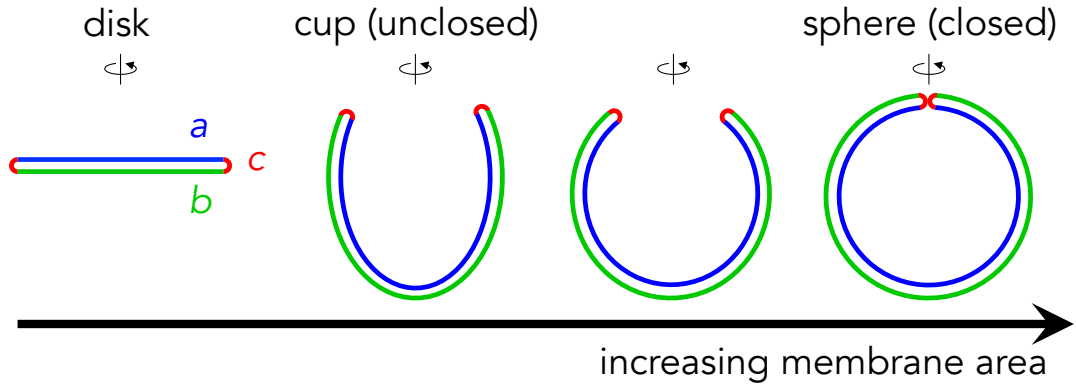

**Figure 1. Schematic representation of autophagosome formation.**
Each shape represents an axially symmetric membrane vesicle. As surface area increases, a phagophore proceeds to transition in its shape from an initial disk to a cup-shaped intermediate to a spherical stomatocyte. The regions *a*, *b*, and *c* indicate the inner membrane, the outer membrane, and the rim, respectively.

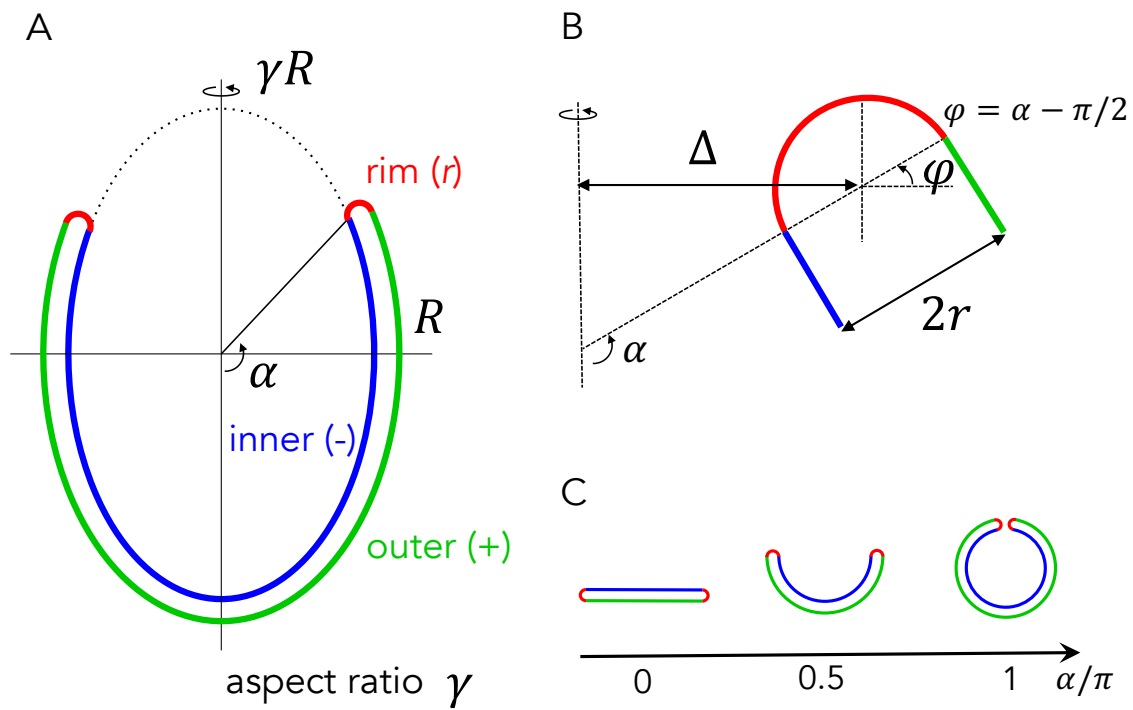

**Figure 2. Description of the phagophore geometry model.**
(*A*) Membrane morphology. (*B*) Enlarged view of the rim and the area around it. (*C*) Correspondence of the bending angle and the morphology.

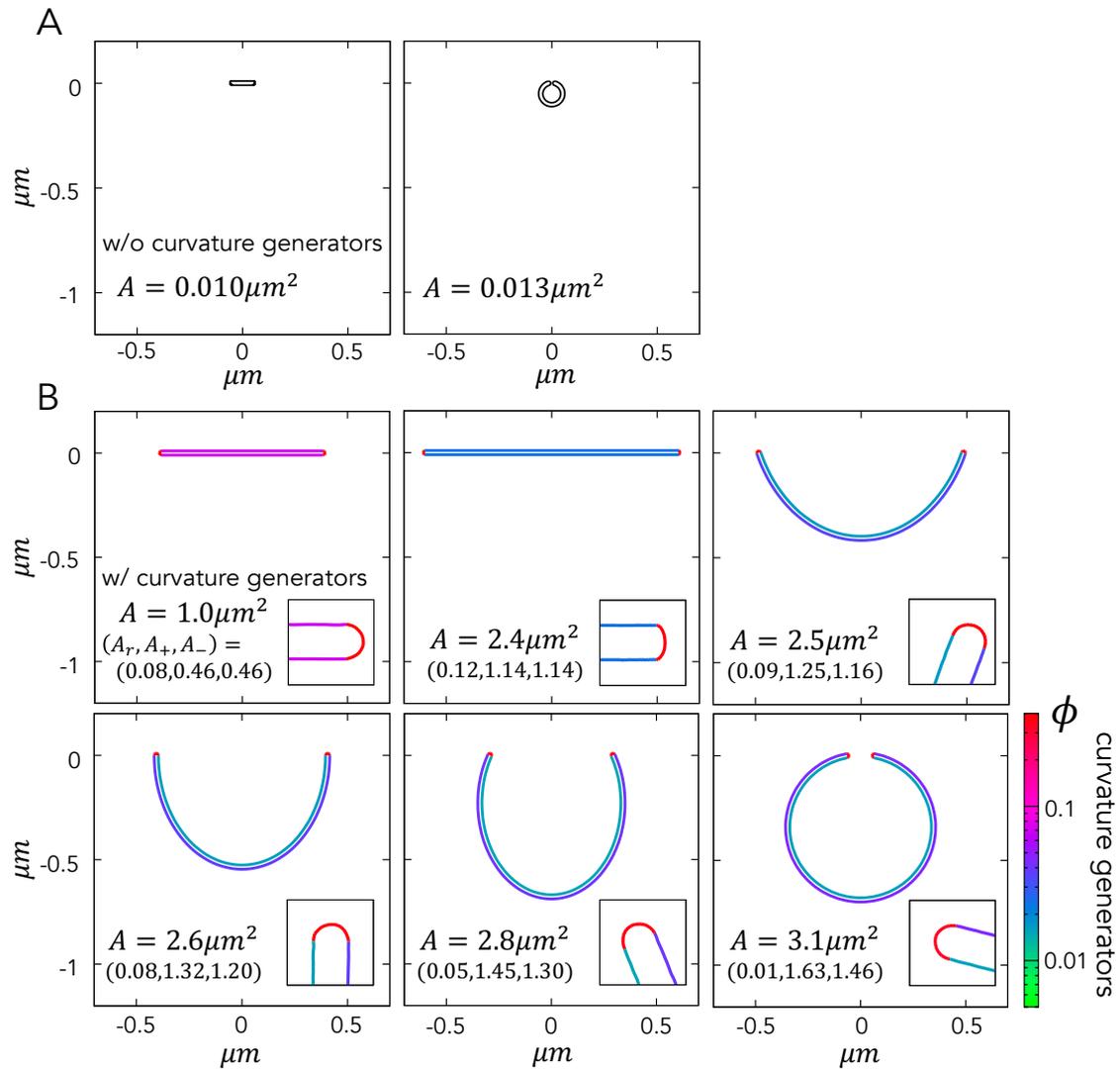

**Figure 3. Progression of the morphological changes as membrane area $A$ increases.** The case with (A) no curvature generator and (B) a fixed area of curvature generators (i.e., abundance) with $A_\phi = 0.1\ \mu m^2$. In panel (B), the color represents the area fraction of curvature generators on a log scale. Each inset depicts an enlarged view of the rim.

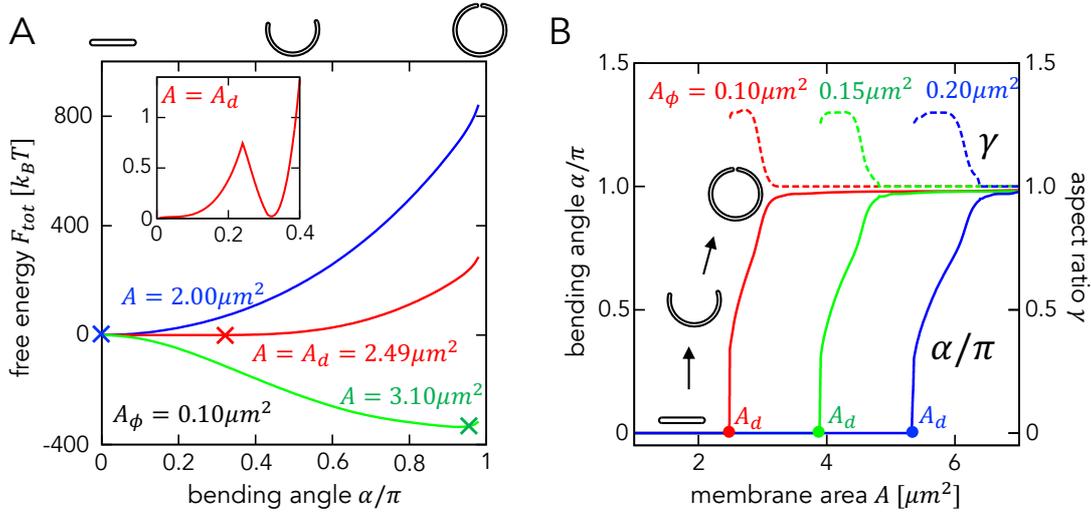

**Figure 4. Relationship between the bending angle and the total membrane area.**
(A) The total free energy as a function of the bending angle $\alpha$ for the membrane area $A = 2.00, 2.49(A_d)$, and $3.10\ \mu m^2$ with a fixed abundance of curvature generators $A_\phi = 0.1\ \mu m^2$. The other internal variables, aspect ratio $\gamma$ and area fraction $\phi_{\pm,r}$, are taken to minimize the free energy at each $\alpha$. The minima are marked by × symbols. The inset shows an enlarged view of the free energy at $A = A_d$. (B) The bending angle (solid) and the aspect ratio (dashed) as a function of the membrane area $A$ for $A_\phi = 0.1$ (red), 0.15 (green), and $0.2\ \mu m^2$ (blue). The transition points ($A_d$) are marked by ● symbols.

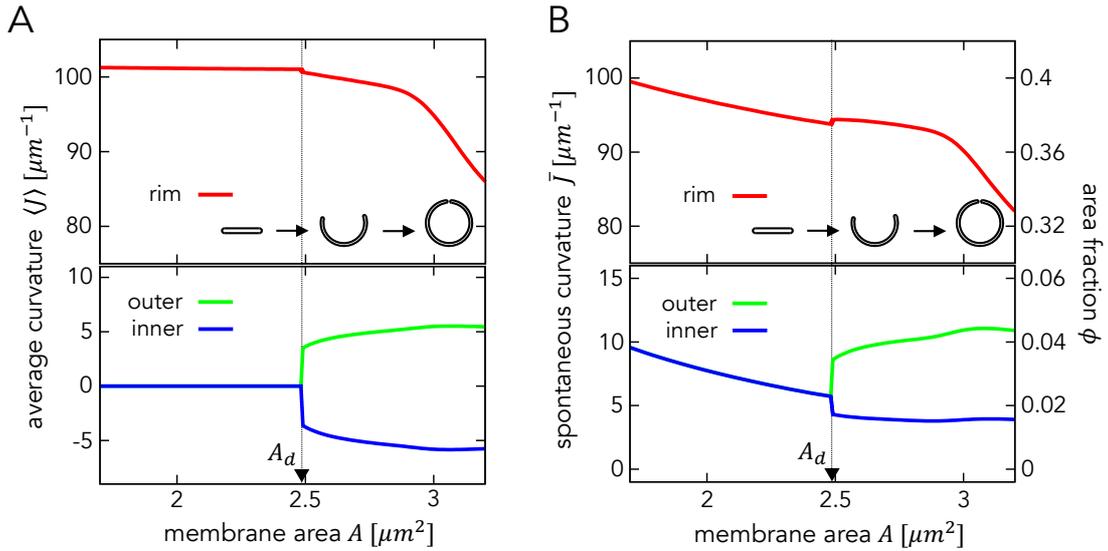

**Figure 5. Relationship between the membrane curvature of each the three parts (i.e., the inner membrane, outer membrane, and rim) and the total membrane area.**

(*A*) Average membrane total curvature and (*B*) spontaneous curvature generated by curvature generators as a function of the membrane area $A$ at $A_\phi = 0.1\ \mu m^2$. Each vertical dotted line represents the transition point from the disk to the cup shape.

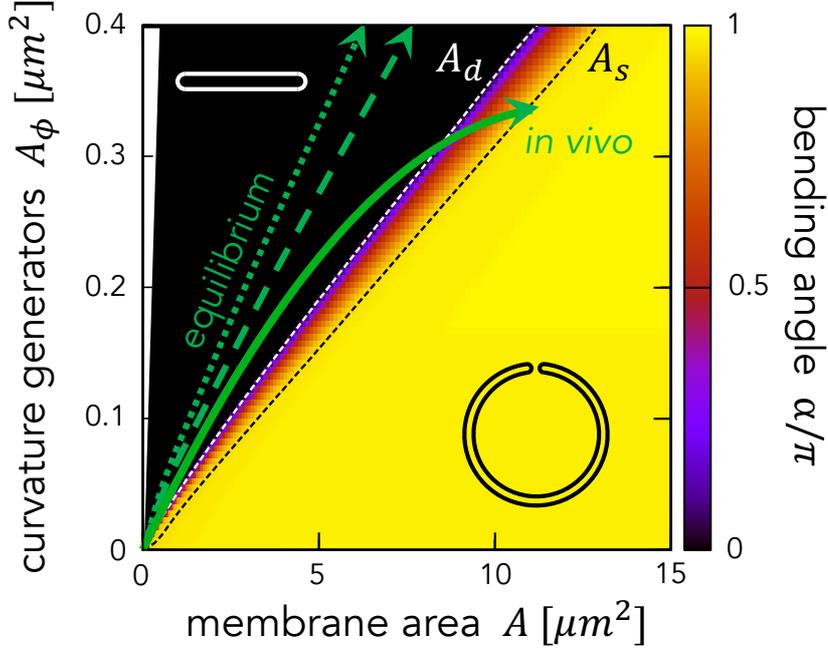

**Figure 6. Heat map of the bending angle on the plane defined by the area (i.e., abundance) of curvature generators, $A_\phi$, and total membrane area, $A$.**
The region $A_\phi > A$ is shown in white. The white and black dashed lines show the boundary $A_d$ (between disk and cup shapes) and $A_s$ (between cup and closed shapes). Changes in the abundance of the curvature generators $A_\phi$ as a function of the membrane area $A$ when the chemical potential is constant, $\mu = \varkappa$, (green dotted line) and proportional to the membrane curvature, $\mu = \varkappa J$, (green dashed line) are shown. The *in vivo* curve is determined by fitting the experimental data (see Figure S6).

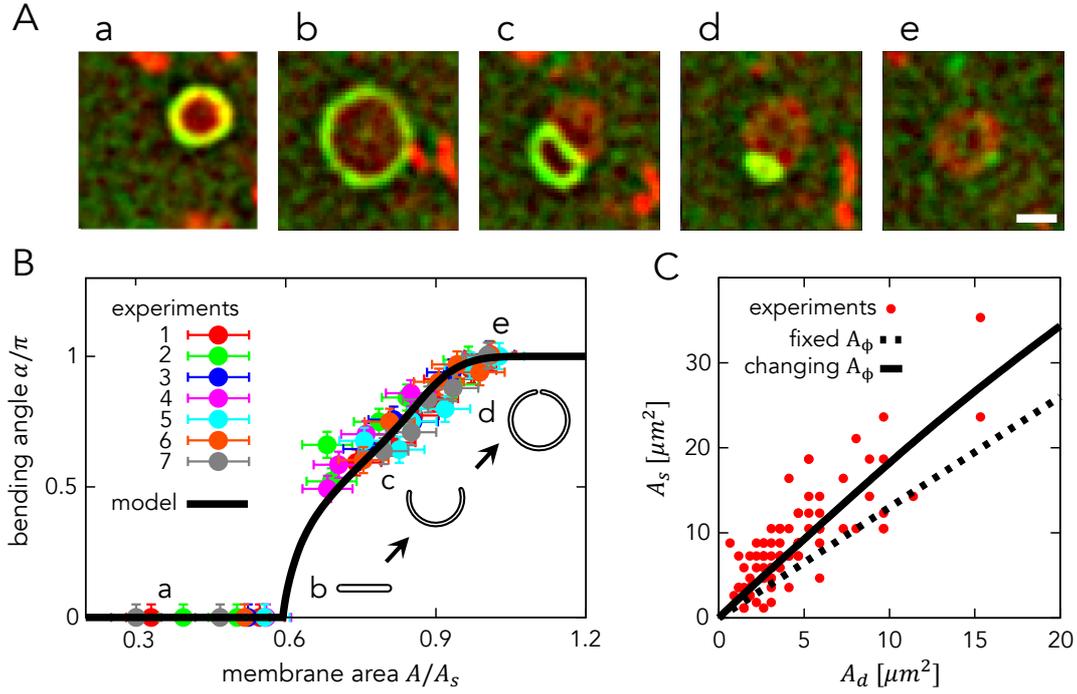

**Figure 7. Comparison of the model predictions with *in vivo* experiments.**
(*A*) Time-lapse imaging of mouse embryonic fibroblasts expressing mRuby3-LC3B and GFP-ATG2A under starvation conditions; scale bar, 1 μm. Video S1 shows a video of this process. (*B*) The bending angle $\alpha$ as a function of the membrane area $A$ normalized by the closed area $A_s$. The solid curve shows the model result in which the abundance of curvature generators changes according to the fitting $\bar{A}_\phi$. Images *a–e* in (*A*) were taken at points *a–e* in experiment 1 in (*B*). (*C*) Relationship between the maximum disk area $A_d$ and the closed spherical area $A_s$ during autophagosome formation. The experimental data were obtained from 89 autophagic structures in starved mouse embryonic fibroblast (MEFs; red dots). The results of the models are shown as dotted and solid lines where the abundance of curvature generator is fixed and changes according to the fitting $\bar{A}_\emptyset$, respectively.

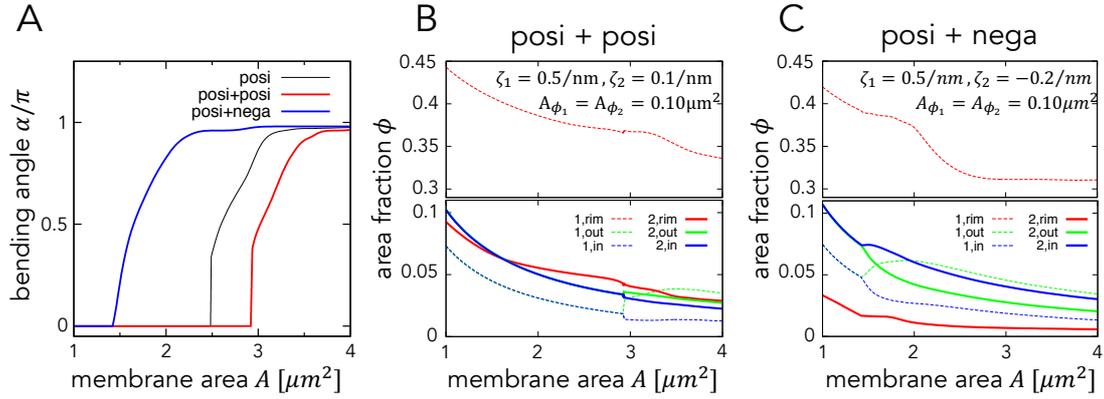

**Figure 8. Effects of two types of curvature generators with spontaneous curvatures $\zeta_1$ and $\zeta_2$.**
(*A*) The bending angle $\alpha$ as a function of the membrane area $A$ for three cases: only one type of positive curvature generator, $\zeta_1 = 0.5\ nm^{-1}$, $A_{\phi_1} = 0.10\ \mu m^2$ (also shown in Figure 4*B*) (posi); positive and weakly positive curvature generators combined, $\zeta_1 = 5\zeta_2 = 0.5\ nm^{-1}$ and $A_{\phi_1} = A_{\phi_2} = 0.10\ \mu m^2$ (posi + posi); and positive and negative curvature generators, $\zeta_1 = -2.5\zeta_2 = 0.5\ nm^{-1}$ and $A_{\phi_1} = A_{\phi_2} = 0.10\ \mu m^2$ (posi + nega). The area fraction $\phi_{i,j}$ of the type-*j* curvature generators in area *i* is shown for (*B*) the posi + posi case and (*C*) the posi + nega case.

**Video S1. Time-lapse imaging of mRuby3-LC3B (red) and GFP-ATG2A (green), related to Figure 7.**
Representative time frame images are shown in Figure 7A; scale bar, 1 μm.

**Table 1.** The parameter values of the model. The parameters, $r$, $\kappa$ and $\zeta$ are taken from (Nguyen et al., 2017), (Helfrich, 1973), and (Campelo et al., 2008), respectively.

| membrane | value | curvature generators | value |
| --- | --- | --- | --- |
| rim radius $r$ | $10\ nm$ | spontaneous curvature $\zeta$ | $0.5\ nm^{-1}$ |
| rigidity $\kappa$ | $20 k_B T$ | effective size $a_\phi$ | $50\ nm^2$ |

## Supplemental Information

**Transparent Methods**

**Geometrical parameters of the model membrane**

The geometry of the outer and inner membranes is modeled as a part of an ellipsoid structure with a bending angle $\alpha$, while the rim geometry is modeled as a part of a torus with a fixed minor radius $r$ (Figure 2). In the ellipsoid approximation, the total curvature $J_+$ ($J_-$) and the surface element $A_+$ ($A_-$) of the outer (inner) membrane is given by

$$J_\pm = \frac{\gamma}{R_\pm} \frac{\cos^2 t + \gamma^2 \sin^2 t + 1}{(\cos^2 t + \gamma^2 \sin^2 t)^{3/2}}, \quad (S1)$$

$$dA_\pm = 2\pi R_\pm^2 \sin t \sqrt{\cos^2 t + \gamma^2 \sin^2 t}\, dt, \quad (S2)$$

where $R_\pm = R \pm r$ is the radius of the short axis of the outer (inner) membrane. The integration variable $t$ runs along the interval $[0, \alpha]$, and $\gamma$ is the aspect ratio of the long axis to the short axis (Figure 2). Note that $\gamma$ is undefined for $\alpha = 0$. In the torus approximation, the total curvature and the surface element of the rim membrane are respectively given by

$$J_r = \frac{1}{r} - \frac{\cos \varphi}{\Delta - r \cos \varphi}, \quad (S3)$$

$$dA_r = 2\pi r (\Delta - r \cos \varphi) d\varphi, \quad (S4)$$

where $\Delta = R \sin \alpha$ is the radius of the aperture (Figure 2). The integration variable $\varphi$ runs along the interval $[\alpha - \pi/2, \alpha + \pi/2]$.

The rim area is $A_r = 2\pi r (\Delta \pi + 2r \cos \alpha)$, and the outer (+) and the inner (−) membrane areas are

$$A_\pm = 2\pi R_\pm^2 F(\gamma, \alpha), \quad (S5)$$

$$F(\gamma, \alpha) = \int_{\cos \alpha}^{1} \sqrt{\gamma^2 + (1 - \gamma^2) x^2}\, dx, \quad (S6)$$

where $F(\gamma, \alpha)$ is the form factor of the ellipsoid.

**Equilibrium conditions**

We consider the membrane morphology and the distribution of curvature generators, which are obtained by minimizing the free energy, Equation (1), for a given total membrane area, $A$, and the abundance of the curvature generators, $A_\phi$. The optimal membrane radius $R^*$, the bending angle $\alpha^*$, the aspect ratio $\gamma^*$, and the area fraction of the curvature generators at each region $(\phi_r^*, \phi_+^*, \phi_-^*)$ are obtained by minimization,

$$F_{tot}(\{X_i^*\}) = min\{F_{tot}(\{X_i\})\}, \quad (S7)$$

with $X_i = (\alpha, \gamma, R, \phi_r, \phi_+, \phi_-)$. Theses parameters satisfy the constraints

$$A_r + A_+ + A_- = A, \quad A_r\phi_r + A_+\phi_+ + A_-\phi_- = A_\phi. \quad (S8)$$

If curvature generators on the membrane can be exchanged with those in the surrounding environment (e.g., cytosol), the equilibrium state is realized by minimizing the grand potential, Equation (5), for a given total membrane area $A$,

$$\Omega(\{X_i^*\}) = min\{\Omega(\{X_i\})\}, \quad (S9)$$

with $X_i = (\alpha, \gamma, R, \phi_r, \phi_+, \phi_-)$. The abundance of curvature generators, $A_\phi$, is variable in this case.

**Effects of two different types of curvature generators**

In the presence of a second type of curvature generator with a different spontaneous curvature $\zeta_j$, the spontaneous curvature of the bending energy and the partitioning entropic energy in the model are slightly modified. The spontaneous curvature, Equation (3), is modified into

$$\bar{J} = \sum_j \frac{1}{2}\zeta_j \phi_{i,j}, \quad (S10)$$

where $\phi_{i,j}$ is the area fraction of type-$j$ curvature generators in area $i$. The partitioning entropic energy, Equation (4), is modified into

$$F_{part} = -k_B T \sum_{i=\pm,r} \left( \sum_j \phi_{i,j} ln\phi_{i,j} + \left(1 - \sum_j \phi_{i,j}\right) ln\left(1 - \sum_j \phi_{i,j}\right) \right) \frac{A_i}{a_\phi}. \quad (S11)$$

Two types of curvature generators with different spontaneous curvatures $\zeta_1$ and $\zeta_2$ are considered for simplicity. Extension to more types of curvature generators is straightforward.

**Evaluation of the membrane area and the bending angle from *in vivo* experiments**

The membrane area and the bending angle were calculated from the image of mRubby3-LC3B and GFP-ATG2A. The contour of membranes was extracted from the region labeled with mRubby3-LC3B. We added a width of 0.2 μm to the contour, which came from the diffraction limit (red in Figure S7). We fitted the data point with an ellipse

$$\left(\frac{(x - x_0)\cos\theta + (y - y_0)\sin\theta}{R}\right)^2 + \left(\frac{(x - x_0)\sin\theta + (y - y_0)\cos\theta}{\gamma R}\right)^2 = 1 \quad (S12)$$

and obtained the radius $R$, the aspect ratio $\gamma$, the origin $(x_0, y_0)$, and the orientation $\theta$. The bending angle was obtained from the origin and overlapping region of the mRubby3-

LC3B contour and GFP-ATG2A contour, to which a width of 0.2 μm was added (green in Figure S7).

For the disk shape ($\alpha = 0$), the membrane area is given by

$$A = 2\pi R^2 + 2\pi^2 rR + 4\pi r^2, \tag{S13}$$

with the rim radius $r = 10\ nm$. By fitting the shape labeled with mRubby3-LC3B with a sphere, the radius $R$ was obtained, and then the area $A$ was obtained. For the cup shape ($\alpha > 0$), the membrane area is given by

$$A = 4\pi(R^2 + r^2)F(\gamma, \alpha) + 2\pi r(\pi R \sin \alpha + 2r \cos \alpha), \tag{S14}$$

$$F(\gamma, \alpha) = \int_{\cos \alpha}^{1} \sqrt{\gamma^2 + (1-\gamma^2)x^2}\, dx. \tag{S15}$$

By fitting the shape of the area labeled with mRubby3-LC3B with a part of an ellipsoid, the bending angle $\alpha$, the radius $R$, and the aspect ratio $\gamma$ were obtained, and then, the area $A$ was obtained.

**Plasmids**

Full-length cDNA of rat microtubule-associated protein 1 light chain 3B (LC3B, GenBank: NP_074058) was subcloned into the pMRX-IP vector (Saitoh et al., 2003), which was generated from pMXs (Kitamura et al., 2003), together with DNA encoding codon-optimized mRuby3 (modified from pKanCMV-mClover3-mRuby3; 74252: Addgene) (Matsui et al., 2018). The pMRX-IP-GFP-ATG2A vector was previously described (Velikkakath et al., 2012).

**Cell culture**

Mouse embryonic fibroblasts (MEFs) were cultured in Dulbecco's modified Eagle's medium (DMEM: D6546, Sigma-Aldrich) supplemented with 10% fetal bovine serum (172012, Sigma-Aldrich), and 2 mM L-glutamine (25030-081, Gibco) in a 5% $CO_2$ incubator. For the starvation treatment, cells were washed twice and incubated in amino acid-free DMEM (048-33575, Wako Pure Chemical Industries) without serum.

**Retroviral preparation and establishment of stable cell lines**

Stable cell lines were generated by a retrovirus-mediated transformation method as previously described (Nishimura et al., 2013).

**Fluorescence microscopy**

Cells stably expressing mRuby3-LC3B and GFP-ATG2A were subjected to live-cell fluorescence imaging using the DeltaVison Elite microscope (GE Healthcare Life

Science) equipped with a PLAPON 60XO oil-immersion objective lens (NA 1.42, Olympus) and a cooled-CCD camera (CoolSNAP HQ2, Photometrics). During live-cell imaging, the dish was mounted in a chamber (INUB-ONI-F2, TOKAI HIT) to maintain the incubation conditions at 37ºC and 5% $CO_2$. Images were acquired at intervals of 10 s for 30 min.

**Supplemental Figures**

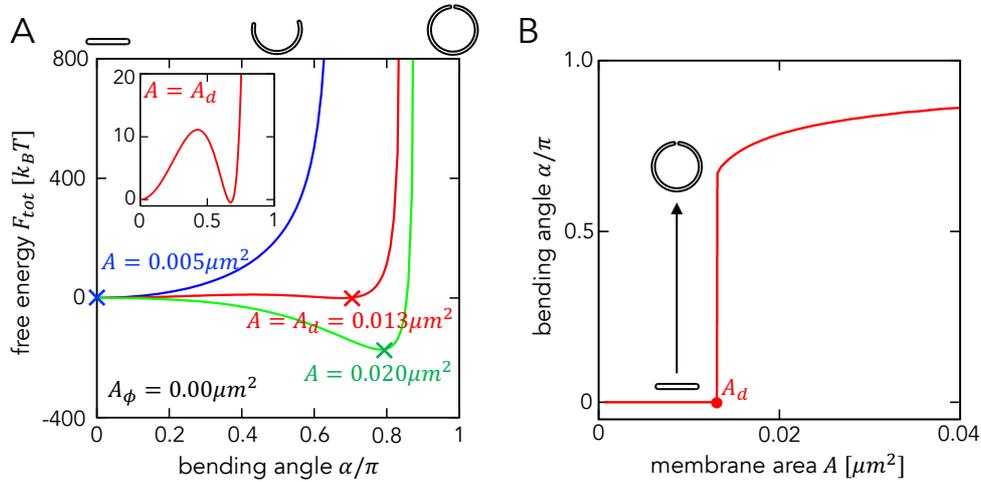

**Figure S1. Relationship between the bending angle and the total membrane area with no curvature generators, related to Figure 4.** (A) The total free energy as a function of the bending angle $\alpha$ for the membrane area $A = 0.005, 0.013\ (A_d)$, and $0.020\ \mu m^2$, where the aspect ratio $\gamma$ is taken to minimize the free energy at each $\alpha$. The minima are marked by × symbols. The inset shows the enlarged view of the free energy at $A = A_d$. (B) The bending angle $\alpha$ as a function of the membrane area $A$. The transition points $(A_d)$ are marked by ● symbols.

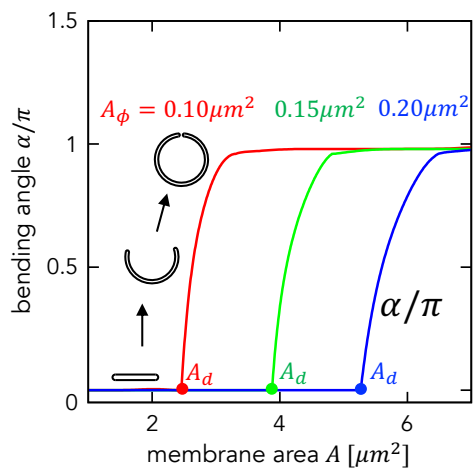

**Figure S2. The bending angle in spherical geometry with a fixed aspect ratio $\gamma = 1$, related to Figure 4.** The bending angle in spherical geometry with a fixed aspect ratio $\gamma = 1$. The bending angle as a function of the membrane area $A$ for $A_\phi = 0.1$ (red), $0.15$ (green), and $0.2\ \mu m^2$ (blue). The transition points $(A_d)$ are marked by ● symbols.

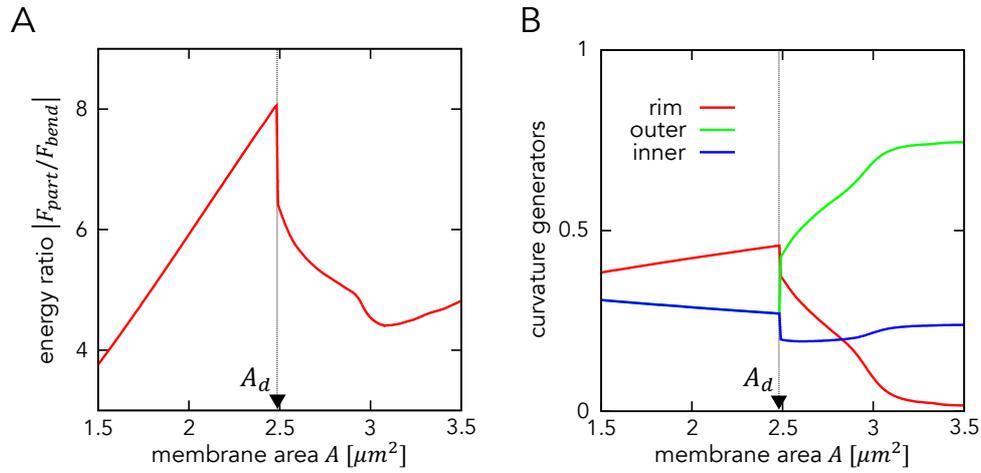

**Figure S3. Relationship between the energy ratio and curvature generator membrane area, related to Figure 5.** (A) Ratio of the partitioning entropic energy $F_{part}$ over the bending energy $F_{bend}$ as a function of the membrane area $A$ for $A_\phi = 0.1\ \mu m^2$. (B) Ratio of curvature generator abundance in each region. The total abundance is fixed at $A_\phi = 0.1\ \mu m^2$.

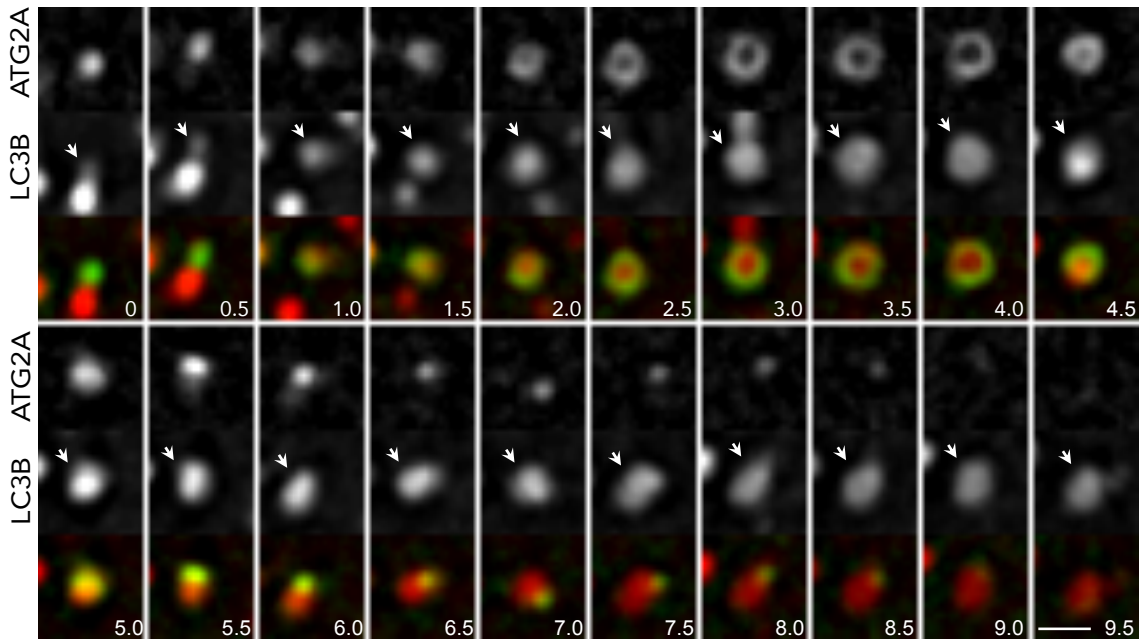

**Figure S4. Time-lapse imaging of autophagosome formation, related to Figure 7.** Mouse embryonic fibroblasts (MEFs) expressing mRuby3-LC3B (red) and GFP-ATG2A (green) were starved, and images were captured every 30 s; scale bar, 1 μm.

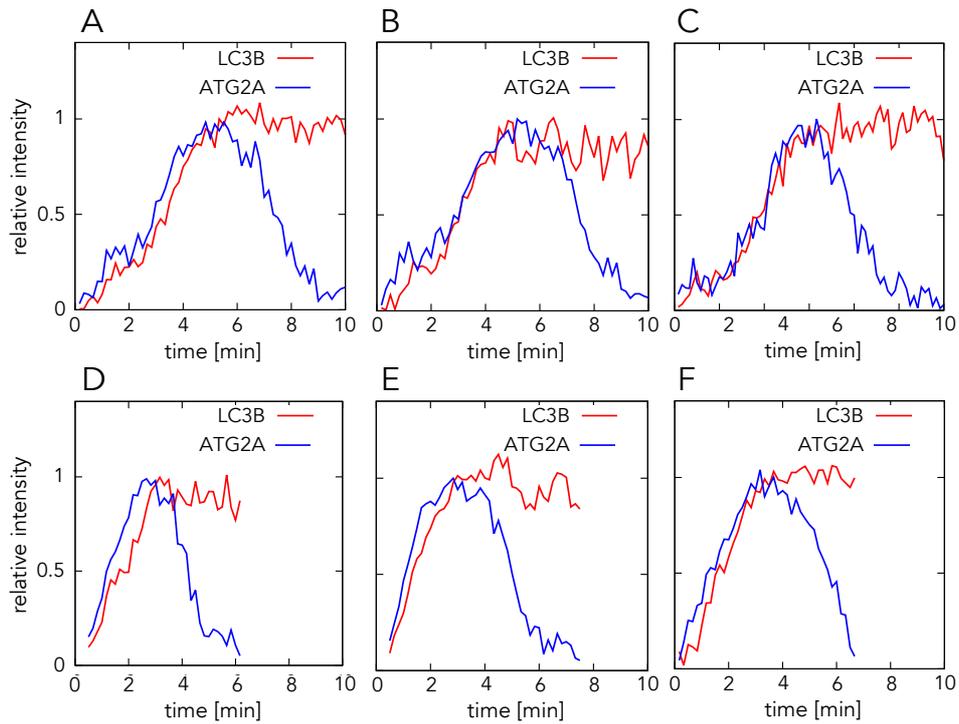

**Figure S5. Quantification of LC3B and ATG2A intensities during autophagosome formation, related to Figure 7.** The time course of the total fluorescent intensity of mRuby3-LC3B (red) and GFP-ATG2A (blue) of each structure is shown throughout autophagosome formation in starved mouse embryonic fibroblasts (MEFs; shown as percentage of maximum intensity). Six independent cases are shown in A–F. Panel A shows the results in Figure S2.

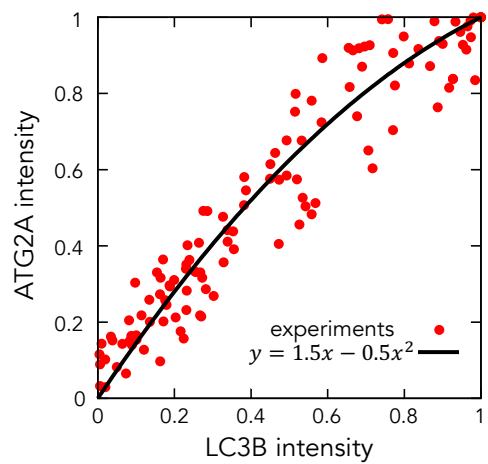

**Figure S6. Correlation between total mRuby3-LC3B and GFP-ATG2A intensities of each structure during the initial phase of autophagosome formation (until their intensities reach a plateau or peak, related to Figure 7.** Each data point is taken from

the results shown in Figure S3. The solid line indicates the second-order polynomial fit of the data.

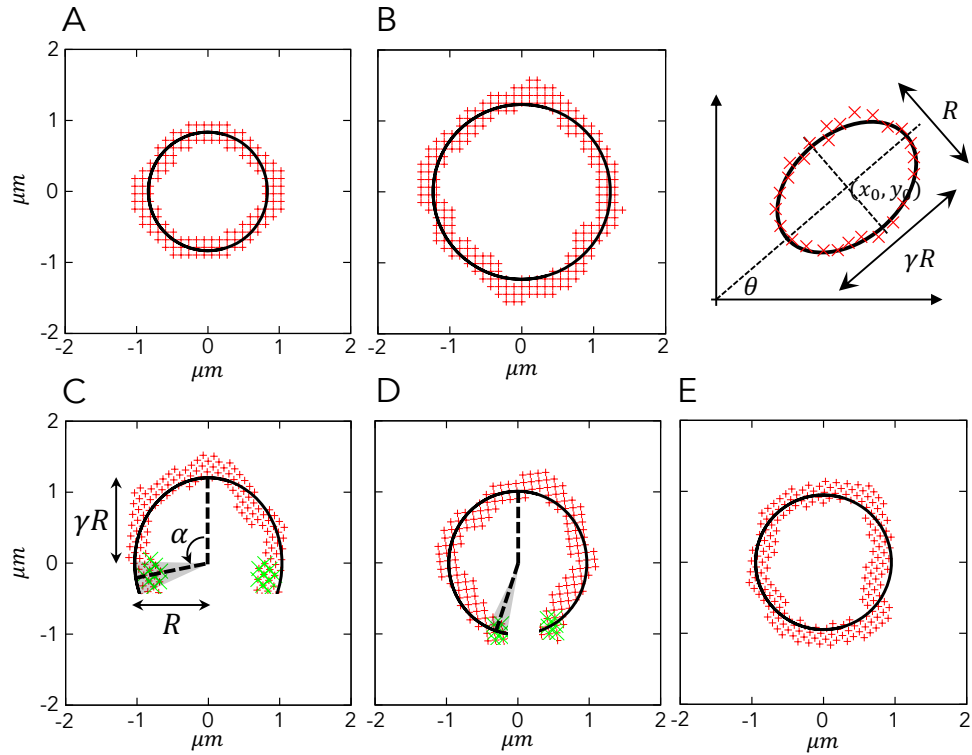

**Figure S7. An example of morphological changes during autophagosome formation, related to Figure 7.** The images in Figure 7A were fitted with a part of an ellipsoid, where the axis dimension is measured in μm. Here the membrane is assumed to take a disk shape in A and B and a cup shape in C–E.